# Strategic Inventories in a Supply Chain with Downstream Cournot Duopoly

Xiaowei Hu*, Jaejin Jang, Nabeel Hamoud, Amirsaman Bajgiran

Industrial and Manufacturing Engineering Department,
College of Engineering and Applied Science,
University of Wisconsin-Milwaukee,
3200 N Cramer St., Milwaukee,
WI 53211, USA

*Corresponding author
Email: hu8@uwalumni.com

**Abstract**

The inventories carried in a supply chain as a strategic tool to influence the competing firms are considered to be strategic inventories (SI). We present a two-period game-theoretic supply chain model, in which a singular manufacturer supplies products to a pair of identical Cournot duopolistic retailers. We show that the SI carried by the retailers under dynamic contract is Pareto-dominating for the manufacturer, retailers, consumers, the channel, and society as well. We also find that the retailer's SI, however, can be eliminated when the manufacturer commits wholesale contract or inventory holding cost is too high. In comparing the cases with and without downstream competition, we also show that the downstream Cournot duopoly undermines the profits for the retailers, but benefits all others.

## 1. Introduction

### 1.1. Background

In the early 1990s, business organisations began to embrace the notion of supply chain management. Studies suggest that effective supply chain management would grant its entities competitive advantage and performance improvement (Li et al., 2006). The study of inventory in a supply chain places great emphasis on cutting down the inventory cost through supply chain coordination, considering inventory is traditionally a "mismatch" between demand and supply (Chopra and Meindl, 2016). Inheriting such concept, a wealth of inventory models (Dey et al., 2016; Krishnamoorthi and Choudri, 2018; Mahapatra et al., 2017; Panda, 2016; Rashid et al., 2018; Tibrewala et al., 2018) are developed to reduce inventory level by seeking the economic order quantity (EOQ).

Generally, inventory is carried in a supply chain for various reasons: to improve service level, to reduce overall logistics costs, to cope with uncertainty in customer demand and lead times, to make seasonal products available throughout the year, to hedge price increase, or to overcome inefficiencies in managing the logistics systems (Ghiani et al., 2004). Inventory can also be carried for reasons other than the aforementioned, namely, to influence the decisions of the market counterparts. This type of inventory is considered to be strategic.

Strategic inventory remained tangential to supply chain studies until 2008 (Anand et al., 2008). In their work, a supplier attempts to maximise his own profit by choosing the wholesale price, whereas a retailer in return attempts to maximise his own profit by choosing the purchasing quantity. The models allow the retailer to hold some of the quantity purchased in one period to be sold in the next period. The authors show that a retailer's strategic inventory mitigates the impact of the supplier's position as a leader. Their work became a platform of much literature reviewed below.

### 1.2. Literature Review

Research in strategic inventory has been well-established. Kirman and Sobel (1974) introduce an intertemporal-dependence decision (e.g. inventory) to the classical static model to form a dynamic model. In a dynamic model, a firm's decisions such as selling quantities or prices vary over time, whilst they do not in a static model. Carrying inventory from one period to the next introduces such dynamicity to the otherwise static models. Aryan and Moses (1982) compare a static model with a dynamic model such as running production at its maximum capacity in one period and shutting down in the following period until inventory is depleted. They find that consumers are better off in the dynamic case when the marginal revenue curve intersects the marginal cost curve at the decreasing portion of the marginal cost curve.

Strategic inventory can be used to threaten potential entrants and deter them from entering the market. Arvan (1985) examines the existence of a subgame-perfect equilibrium of two competing firms that produce and sell in the first-period, and carry the unsold output as inventory for the second period. In the second period, the firms decide whether to produce more to sell. The author shows that even when revenues are symmetric, the equilibrium, if exists, is not symmetric; and that one firm must act as a leader in the second period. The author also shows that in the case of a firm facing a threat from the possibility of entrants, there exists an equilibrium in which one firm accumulates large inventory to deter the entry of potential competitors. Ware (1985) models an incumbent and a potential entrant, where the incumbent has the option of carrying inventory. The author demonstrates that market entrance can be deterred by the threat of carrying inventory. He also argues that if a capacity investment is combined with inventory investment, it would have a greater impact on entry deterrence.

Mitraille and Moreaux (2013) study the existence of Nash equilibrium in a pure strategy of a multiple Cournot firm competition. It was found that the firms carrying inventory act as leaders and gain a larger market share. Strategic inventory also allows a firm to raise its next period sales (Mollgaard et al., 2000). The authors show that the strategic value of inventories depends on the convexity of the cost function, the slope of each firm's demand function, and the cost of storage. Amihud and Medenelson (1989) show empirically that inventory plays a significant role in smoothing the market price in the events of supply and demand shocks and that the firms with more market power tend to hold more inventory.

Strategic inventory can also be utilized to induce cooperation among firms. Matsumura (1999) finds that, in a more-than-two-period model, the firms tend to behave collaboratively. If a firm deviates from the collusion, the non-deviating firm acts as a Stackelberg leader, and the deviating firm acts as a Stackelberg follower. Rotemberg and Saloner (1989) show that firms are tempted to deviate when demand is high. In their empirical study, the authors show that inventory can be used to deter a cheating firm from deviating from a collusive agreement.

Anand et al. (2008) are the first to introduce strategic inventory to a supply chain. They study the interaction between an upstream supplier and a downstream retailer. The retailer decides the purchasing quantity after the supplier announces the wholesale price. They show that the retailer's threat of holding inventory induces the supplier to increase the wholesale price of the first period. A number of studies, thereafter, confirm and extend the findings of Anand et al. Hartwig et al. (2015) empirically study the impact of strategic inventory on a supply chain, showing that the inventory induces the supplier to raise her first period wholesale price. They also find that strategic inventory could enhance the supply chain's performance. Desai et al. (2010), introduce competition to the supply chain model, and examine retailer's forward buying and holding inventory at a cost in more complex competitive channel structures by focusing on marketing variables of merchandising support and trade promotions. The authors find that in such competitive environments, there are situations in which retailers engage in forward buying because of competitive pressures in a prisoner's-dilemma situation.

Zhang et al. (2010) study the impact of asymmetric information of inventory on the optimal contract. In their work, the demand information is known to both supplier and retailer, whereas the sales history is known to the retailer only. The decisions the supplier and the retailer face are the type of contract and the order quantity, respectively. The study shows that the more the retailer stockpiles, the less attracted the supplier to trade. Overall, asymmetric information makes a negative impact on supply chain efficiency.

The above studies assume that the supplier's output is bounded by the market demand. However, in many industries, the supplier's output is also limited for reasons such as production and space capacities. Keskinocak et al. (2008) analyze a two-period model with a manufacturer's limited capacity. They show, when the capacity is under a critical level, the inventory loses its strategic value and the equilibria of the dynamic and commitment contracts become the same.

As shown above, the retailer's strategic inventory forces the supplier to raise his first-period price in response to the retailer's action. However, there are other strategies to respond to the retailer's action such as offering a direct rebate to consumers. Arya and Mittendorf (2013) study the effect of the manufacturer's rebate on the strategic inventory held by the retailer. It is found that, by offering a rebate to consumers, the manufacturer can mitigate the retailer's advantage of holding inventory. They also test the effect of such rebate on a three-period model and find that the manufacturer prefers to offer a limited-time rebate over a non-expiring rebate.

In Arya et al. (2015), a retailer procures a product for multiple divisions, each of which serves a different market. A challenge the retailer faces is whether to centralize or decentralize its inventory. The authors find that when the retailer decentralizes inventory, the moves of the manufacturer (i.e. raising the price of the 1st period) and the retailer (i.e. carrying more inventory) are less aggressive.

The studies reviewed above largely assume that the product under consideration is homogenous; however, quality deterioration is often important. Saloner (1986) examines a dynamic model of two firms carrying inventory to the second period assuming that the value of inventory decreases over time due to product obsolescence. The author shows that the cost of such obsolescence induces firms not to carry inventory.

Mantin and Jiang (2017) study the effect of quality deterioration on strategic inventory. They observe that when quality deteriorates over time, the wholesale price and the retail price in the first period are not always higher than the second period. They show that the retailer is better off when the deterioration rate is high, while the supplier is better off when the rate is low. They also show that the profits of the supplier, the retailer, and the channel depend on the level of deterioration and the holding cost. Under certain thresholds, all agents of the supply chain would benefit from carrying strategic inventory.

In Antoniou and Fiocco (2017), not only the retailer is allowed to carry inventory, but also the supplier. The supplier holds inventory in response to the retailer's stockpiling in the first period. The authors show that the response of the retailer is similar to that of the commitment contract, where the retailer's stockpiling is eliminated. Viswanathan (2016) studies downstream competition between retailers assuming that the level of inventory of both retailers is the same. The author compares the effect of downstream competition under two different contracts: dynamic and commitment contracts. The analysis shows that under the dynamic contract, the downstream competition decreases the manufacturer's profit, the retailers' profits, and the consumer surplus. However, under the commitment contract, only the manufacturer is better off when there is downstream competition.

Yet, to this end, no work has examined the circumstances of a supply chain in which carrying stocks would be motivated and/or beneficial. Nor has any previous work explicitly linked SI to its holding cost, or to the market power of the subjects that may invoke SI in a competitive setting. To fill this gap, in this paper, we consider a vertical supply chain, in which a singular manufacturer supplies products to a pair of identical retailers in Cournot-duopolistic competition. We focus on the game-theoretic aspect of the behaviors of the manufacturer and suppliers. Our central insight is that allowing inventory remains an effective strategy in a retailer duopoly for not only the retailers, but also the manufacturer, consumers, and the entire society; it benefits all parties' welfare. However, the effectiveness of strategic inventory appears to be vulnerable in two aspects. First, too high of a holding cost will discourage the retailers to carry inventories. Second, the manufacturer's employment of a commitment wholesale contract could eliminate the retailer's strategic inventory, although such an outcome would result in all parties in the supply chain being worse off. We explain the retailer's vulnerability by means of the Lerner Index in a static market. Moreover, we further discover that, compared to a retailer monopoly, the Cournot-like competition between the downstream retailers undermines the aggregated profit of retailers even when they opt to carry inventory, despite that such competition benefits all other parties, the channel, and society.

The rest of this paper is organised as follows. Section 2 describes the problem and presents the models. Section 3 discusses the result of the equilibrium analysis. Section 4 highlights the key findings and concludes the paper.

## 2. The Model

We begin by constructing a supply chain model, in which a singular manufacturer attempts to sell his products to a market via two retailers. The manufacturer determines the wholesale price of the product and possesses complete bargaining power over the retailers. The two retailers compete with one another as Cournot duopolists, i.e., their choice of output quantities influences the market price. We stipulate the manufacturer to be upstream and the retailers to be downstream in this vertical supply chain. Two time periods are considered. We allow the retailers to store goods from one period to the next in the form of inventories, and assume the goods are non-perishable. We study the effect of inventories by examining the economic measures of the manufacturer, retailers, consumers, the supply chain, and society.

In the initial model, the economic activities manifest as follows. At the beginning of period 1, the manufacturer announces a linear contract with a wholesale price of $w_1$ per unit to the retailers, assuming all production cost is normalized to 0. Upon observing this information, retailer $j$ $(j = 1, 2)$ determines and purchases the product of quantity $q_{1j} + I_j$ from the manufacturer, selling $q_{1j}$ to the market, and carrying the remainder $I_j$ as inventory to period two, incurring an aggregated unit inventory holding cost, h per unit per period. We assume such cost is only incurred by the retailers in the period when the inventory is purchased (i.e., period 1). In period two, the manufacturer again announces another linear contract of the identical goods with wholesale price $w_2$ to the retailers. Retailer $j$ orders a quantity of $q_{2j}$, and sells $q_{2j} + I_j$ to the market in entirety. The sequence of event is depicted in Figure 1.

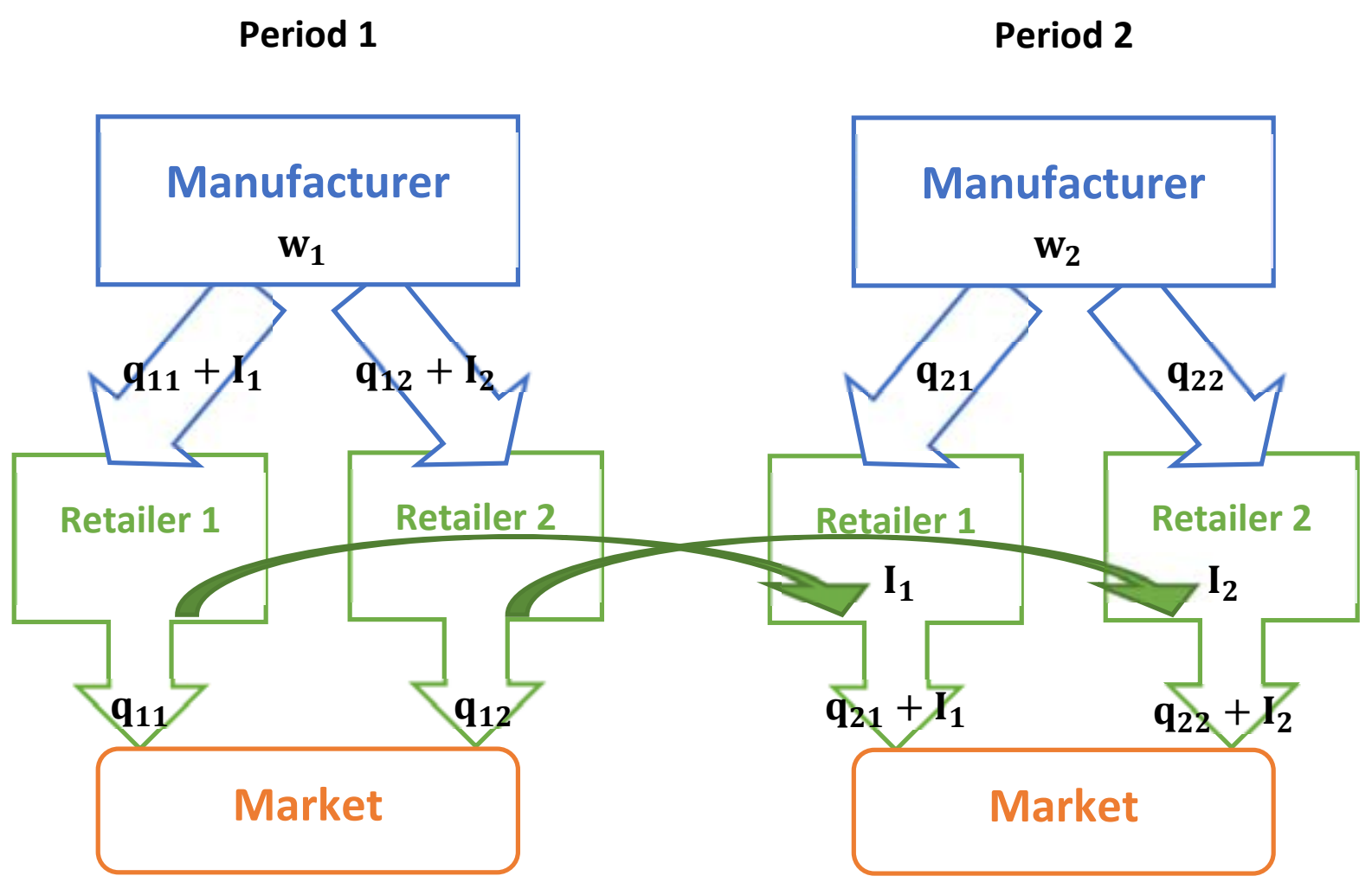

*Figure 1 Strategic inventory in a supply chain with downstream duopoly*

The retail price of each period is determined by a linear inverse price-demand function. That is, the unit price of a product sold in the market is p(Q) = a-bQ, where “a” is the reservation price, an exogenous and strictly positive value. We assume the market demand for the product to be sufficiently large, two retailers to be identical, the interest rate and the depreciation rate of the product in inventory to be reflected in the aggregated unit inventory holding cost rate, h.

Both the manufacturer and the retailers attempt to maximise the profit of their own over the entire time horizon: the first and second periods. In each period, the manufacturer’s decision is his wholesale pricing strategy, whereas each retailer’s decision includes her procurement and inventory quantities. Intuitively, for the manufacturer, a higher wholesale price will directly lead to a higher revenue, but will also discourage retailers’ procurement via suppressing consumer demand. For the retailers, a larger procurement may benefit the total revenues and will certainly benefit the manufacturer, but could also lower the market price and hurt the retailers’ own profit. Moreover, an appropriate inventory will serve in the retailer’s favor if the wholesale price rises, but may also undercut the market price, leading to a decreased profit. Carrying inventory will certainly hurt the manufacturer in the event when the manufacturer intends to raise the wholesale price in the next period. On balance, each firm’s maximum profit would not be solely driven by its own decisions, but also by any other firms’ action comprehensively.

Obviously, the inventory holding cost h is a drain of profit for retailers; if h is too high, holding inventory may hurt the retailer’s profitability and undermine the retailer’s motivation to adopt such strategy. Therefore, we first base our main findings on the assumption of holding cost being $0<h<a/4$. In section 3.4, we provide more discussion on this assumption and explore its impact on each firm’s decisions.

In the following two subsections, we construct a game-theoretical model with two different types of manufacturer wholesale contracts. In section 2.1, we consider an initial model, in which case, the manufacturer employs a dynamic wholesale contract. Under this contract, the wholesale price is revealed only at the beginning of each respective period. In section 2.2, we alter the initial model by considering a commitment wholesale contract, in which case, the manufacturer announces the wholesale prices for both subsequent time periods at the beginning of period 1. The questions we attempt to answer are: what might be each firm's best practice? How sensitive are strategic inventories with respect to the cost of carrying them? Compared to monopoly, how does the Cournot duopoly in downstream impact the decisions and profit of the manufacturer and retailers? Do the entire supply chain and society do better or worse with different wholesale contracts by the manufacturer and the corresponding inventory strategies by the retailers?

**2.1 Dynamic Wholesale Contract**

This subsection examines a dynamic contract case, in which the manufacturer reveals the wholesale contract of each period to the retailers only at the beginning of such period. When the retailers observe the contractual information (i.e., the wholesale price), they proceed to determine their purchase and sales quantities to maximise their profits.

Since all goods will be sold out to the market by the end of period 2, retailer j's profit function in period 2 is

$$\pi R_{2j}(q_{2j}) = [a - b\,(q_{21} + q_{22} + I_1 + I_2)](q_{2j} + I_j) - w_2 q_{2j} \tag{1}$$

and her problem becomes

$$\max_{q_{2j} \geq 0} \pi R_{2j} \tag{2}$$

At the beginning of period 2, the manufacturer's profit is

$$\pi M_2(w_2) = w_2(q_{21} + q_{22}) \tag{3}$$

and his problem becomes

$$\max_{w_2 \geq 0} \pi M_2 \tag{4}$$

In period 1, retailer j's profit is,

$$\pi R_{1j}(q_{1j}, I_j) = [a - b\,(q_{11} + q_{12})]q_{1j} - w_1(q_{1j} + I_j) - hI_j \tag{5}$$

and her problem in period 1 becomes

$$\max_{\{q_{1j} \geq 0, I_j \geq 0\}} \pi R_{1j} + \pi R_{2j} \tag{6}$$

At the beginning of period 1, the manufacturer's profit is

$$\pi M_1(w_1) = w_1(q_{11} + q_{12} + I_1 + I_2) \tag{7}$$

Similarly, the manufacturer's decision problem is

$$\max_{w_1 \geq 0} \pi M_1 + \pi M_2 \tag{8}$$

We solve the problem by backward induction reasoning, acknowledging the equilibrium obtained for the entire stage game is subgame-perfect (Selten, 1975).

Lemma 1 establishes the unique subgame-perfect Nash Equilibrium (SPNE) under the dynamic contract setting.

*Lemma 1*

*When the inventory holding cost is high ($\frac{a}{12} \leq h < \frac{a}{4}$), the unique SPNE in the manufacturer's pricing and retailers' quantity is*[1]

$$\{{}^{D}\widehat{w}_1^*,\ {}^{D}\widehat{w}_2^*\} = \{\frac{a}{2}, \frac{a}{2}\}$$

$$\{{}^{D}\hat{q}_{11}^*,\ {}^{D}\hat{q}_{12}^*,\ {}^{D}\hat{q}_{21}^*,\ {}^{D}\hat{q}_{22}^*,\ {}^{D}\hat{I}_1^*,\ {}^{D}\hat{I}_2^*\} = \{\frac{a}{6b}, \frac{a}{6b}, \frac{a}{6b}, \frac{a}{6b}, 0, 0\}$$

*When the inventory holding cost is low ($h < \frac{a}{12}$), the unique SPNE is*

$$\{{}^{D}w_1^*,\ {}^{D}w_2^*\} = \{\frac{3}{124}(21a - 4h), \frac{1}{124}(55a + 84h)\}$$

$$\{{}^{D}q_{11}^*,\ {}^{D}q_{12}^*,\ {}^{D}q_{21}^*,\ {}^{D}q_{22}^*,\ {}^{D}I_1^*,\ {}^{D}I_2^*\} = \{\frac{61a + 12h}{372b}, \frac{61a + 12h}{372b}, \frac{55a + 84h}{372b}, \frac{55a + 84h}{372b}, \frac{7(a - 12h)}{186b}, \frac{7(a - 12h)}{186b}\}$$

The derivation is included in Appendix A. Lemma 1 characterizes that the SPNE essentially depends upon the retailer's holding cost. We observe that $\{{}^{D}w_1^*, {}^{D}w_2^*\}$ and $\{{}^{D}q_{11}^*, {}^{D}q_{12}^*, {}^{D}q_{21}^*, {}^{D}q_{22}^*, {}^{D}I_1^*, {}^{D}I_2^*\}$ collapse into $\{{}^{D}\widehat{w}_1^*, {}^{D}\widehat{w}_2^*\}$ and $\{{}^{D}\hat{q}_{11}^*, {}^{D}\hat{q}_{12}^*, {}^{D}\hat{q}_{21}^*, {}^{D}\hat{q}_{22}^*, {}^{D}\hat{I}_1^*, {}^{D}\hat{I}_2^*\}$, respectively, when $h = \frac{a}{12}$. We also observe ${}^{D}w_1^*$ is strictly greater than ${}^{D}w_2^*$ when $h < \frac{a}{12}$; that is, when holding cost is low enough, each retailer will still carry inventory even with an anticipated manufacturer's wholesale price declining between periods – a phenomenon of strategic inventory case portrayed by previous studies (Anand et al., 2008; Arya and Mittendorf, 2013).

The intuition primarily lies in the action taken by the market-power-dominating manufacturer on the retailers' response. Let us consider two inventory holding cost levels. When the holding cost is high, the retailer's best practice is to avoid carrying inventory between periods, because the holding cost is a drain of the retailer's profit. Anticipating this action, the manufacturer sets up a uniform wholesale price between two periods, since an unequal inter-period wholesale price will eliminate sales in the high-price period.

When the holding cost is low, clearly each retailer is able to build up inventories needed with less cut to her profit. The retailer is, in fact, motivated to do so, because carrying inventories from period 1 could coerce the manufacturer to price down in period 2 by virtually introducing a "competitor". Thus, a delicate decision left for the manufacturer here is his period-1 wholesale price. Such price needs to be high enough to preclude the retailers from carrying too much inventory, but not too high for himself to become unprofitable. In other words, the rationality for the manufacturer's declining wholesale pricing strategy is that such strategy is the manufacturer's attempt to discourage the retailers' motivation to build up stocks from period 1. In section 3, we justify our claims with respect to the firms' market power and sensitivity to holding cost by analytical evidence.

Conversely, consider the market-power-dominated retailers' response to the manufacturer's price offerings. If the manufacturer offers a uniform or increasing inter-period contract, each retailer in response would relinquish inventory, achieving Just-In-Time (JIT). If the manufacturer offers a declining inter-period contract, each retailer would then choose to hold inventory as a threat to mitigate the market power of the monopolistic manufacturer in period 2. Cleary, such threat cannot be credible unless holding inventory is not too costly. Desai et al. (2010) observe the similar inventory-carrying behavior from the retailers with respect to the holding cost.

[1] In this paper, we use left-superscript D and C to denote dynamic and commitment contract, right-superscript Mon and Duo to denote retailer monopoly and duopoly, and $\frown$ to denote high inventory holding cost, hereafter.

### 2.2 Commitment Wholesale Contract

In this subsection, we explore another scenario, in which the manufacturer employs a commitment contract, i.e., at the onset of the first period, the manufacturer announces the wholesale prices for both subsequent periods. Here, the commitment contract resembles a future contract, a type of contract used by wholesalers in order to hedge the volatility of future market prices.

Similar to the previous model, retailer j's profit in period 2 is given by (1), and her procurement problem is (2). In period 1, retailer j's profit is given by (5), and her procurement and inventory problem becomes (6). The manufacturer's profit in each period is given by (7) and (3), respectively. At the beginning of period 1, the manufacturer shall determine both $w_1$and $w_2$. Hence his production problem becomes

$$\max_{w_1 \geq 0, w_2 \geq 0} \pi M_1 + \pi M_2$$

We again solve the problem by backward induction. Lemma 2 establishes the unique SPNE under the commitment contract.

*Lemma 2*

*When the manufacturer exercises commitment wholesale contract, the unique SPNE in the manufacturer's pricing and retailers' quantity is*

$$\{{}^{C}w_1^*,\ {}^{C}w_1^*\} = \{\frac{a}{2}, \frac{a}{2}\}$$

$$\{{}^{C}q_{11}^*,\ {}^{C}q_{12}^*,\ {}^{C}q_{21}^*,\ {}^{C}q_{22}^*,\ {}^{C}I_1^*,\ {}^{C}I_2^*\} = \{\frac{a}{6b}, \frac{a}{6b}, \frac{a}{6b}, \frac{a}{6b}, 0, 0\}$$

The derivation is included in Appendix B. Lemma 2 characterizes that the manufacturer's commitment contract eliminates retailers' necessity of carrying inventory regardless of the inventory carrying cost, *h*. Such a contract ascertains the retailer's response (JIT) in the sense of putting forward pricing information in exchange for consistent periodic procurements.

## 3. Results and Findings

In this section, we compare the dynamic and commitment contract under a retailer duopoly; we also compare each party's behaviors under such setting against those under retailer monopoly. We focus on performance measures of each firm in the supply chain, and also on the total profit of the supply chain, consumer surplus, and social welfare.

Consumer Surplus (CS) of period $i$ ($i = 1, 2$) is given by

$$CS_i = \frac{1}{2}(a - p_i)q_i ,$$

where $q_i$ and $p_i$ represent the market quantity and price, respectively.

The channel profit of each period is the sum of the manufacturer and retailers' profits, given by

$$\pi_i = \pi M_i + \pi R_{i1} + \pi R_{i2} .$$

Also, we denote the social welfare (SW) as the accrual of CS and channel profit in each period.

$$SW_i = CS_i + \pi_i$$

### 3.1 Comparison of Dynamic vs. Commitment Contracts

In this subsection, we compare the impact of the manufacturer's wholesale contractual strategies on the supply chain.

Proposition 1 establishes a pair of necessary-and-sufficient conditions in order for a given retailer's strategic inventory to be invoked.

*Proposition 1*

*The retailers' strategic inventory exists when and only when: 1) the manufacturer employs a dynamic wholesale contract, and 2) the retailer's inventory carrying cost is less than a/12.*

The first condition stipulates the existence of strategic inventory to the manufacturer's contractual strategy, an exogenous criterion of the retailers, whereas the second condition requires the maximum threshold of the cost in order for holding the inventories to be feasible, an endogenous criterion of the retailers. The proof of Proposition 1 is immediate from Lemma 1 and Lemma 2.

Moreover, because the manufacturer's strategy-set in our framework is between the two types of contracts, an important implication from the first condition is that the manufacturer's commitment contract eliminates the retailer's strategic inventory. Clearly, the commitment contract in this setting is essentially a layout of complete information concerning the wholesale price across the entire time horizon. Such information leads to a uniform wholesale price between periods and obsoletes the inventories carried by retailers, because carrying inventory undoubtedly would result in additional costs and harm the retailers' profitability.

Next, we discuss how strategic inventories impact each supply chain agent's performance. The most interesting case is when the retailers' strategic inventories are triggered. Acknowledging Proposition 1 and Lemma 1, we shall be able to carry out the desired comparison by setting the inventory holding cost to be low, i.e., 0<h<a/12. Then the presence of strategic inventories will solely depend on the wholesale contractual strategies adopted by the manufacturer. Proposition 2 establishes the strategic inventory's advantages delineated in Table 1 in a conclusive manner. Its proof is immediate.

*Proposition 2*

In a Cournot duopoly of retailers, the manufacturer, retailers, and consumers are all better off under a dynamic contract.

Table 1 displays the selected performance comparisons and outcomes between the dynamic contract and commitment contract. The complete results from a more generalized setting are shown in Table 7 of Appendix E.

*Table 1 Retailer Duopoly: Commitment vs. Dynamic Contract under Low Holding Cost*

| | | **Dynamic** | | **Commitment** |
|---|---|---|---|---|
| Wholesale Price | $w_{avg}$ | $\frac{125a^2-24ah+144h^2}{4(65a-36h)}$ | < | $\frac{a}{2}$ |
| Retail Price | $p_{avg}$ | $\frac{7849a^2-1632ah-3600h^2}{186(65a-36h)}$ | < | $\frac{2a}{3}$ |
| Consumer Purchase Qty | $Q_{total}$ | $\frac{65a-36h}{93b}$ | > | $\frac{2a}{3b}$ |
| Inventory | I | $\frac{7(a-12h)}{186b}$ | > | 0 |
| Retailer (ea.) Sales | $q_{total}$ | $\frac{65a-36h}{186b}$ | > | $\frac{a}{3b}$ |
| Retailer (ea.) Profit | $\pi R_{total}$ | $\frac{4073a^2-3636ah+10656h^2}{69192b}$ | > | $\frac{a^2}{18b}$ |
| Manufacturer Profit | $\pi M_{total}$ | $\frac{125a^2-24ah+144h^2}{372b}$ | > | $\frac{a^2}{3b}$ |
| Supply Chain Profit | $\pi_{total}$ | $\frac{7849a^2-2934ah+12024h^2}{17298b}$ | > | $\frac{4a^2}{9b}$ |
| Consumer Surplus | $CS_{total}$ | $\frac{4241a^2-5064ah+3600h^2}{34596b}$ | > | $\frac{a^2}{9b}$ |
| Social Welfare | $SW_{total}$ | $\frac{19939a^2-10932ah+27648h^2}{34596b}$ | > | $\frac{5a^2}{9b}$ |

Several observations can be made. First, we notice that with the dynamic contract, the first (second) period wholesale price is higher (lower) than the respective prices with the commitment contract; i.e., ${}^{D}w_1 > {}^{C}w_1 = {}^{C}w_2 > {}^{D}w_2$. The intuition for the decreasing wholesale price trend under dynamic contract is that, in period 1 the manufacturer is capable of capturing as much profit as possible by pricing the products high, knowing that during this sales season no inventory from the prior period exists. The manufacturer also anticipates with certainty that the retailers will carry inventory for the next period since the wholesale prices are not below the inventory-eliminating level $\frac{a}{12}$ for both periods. Hence, in order to discourage the retailer's credible threat to the manufacturer's monopoly power by stocking up products for the next sale period, the manufacturer must increase the price in period 1 and decrease the price in period 2.

Second, we observe that the average retail price with the dynamic contract is lower than with the commitment contract, i.e., ${}^{D}p_{avg} < {}^{C}p_{avg}$, where the average prices are weighted by the total output (consumer purchase) quantity in each period. Because the manufacturer and each retailer are self-interest agents in a single vertical supply chain system, the product's price has to experience markup twice before reaching the market, resulting in a higher sale price and eroding the consumer surplus. Such a phenomenon is called double marginalization (Spengler, 1950). Nonetheless, in our retailer duopoly framework, the dynamic contract and its resultant strategic inventories could ameliorate the losses due to double marginalization. We hence claim that strategic inventory reduces the double-marginalization effect.

Third, as is evident, we recognize that a dynamic contract is superior to a commitment contract for both the manufacturer and retailers, as well as the entire channel, and society. This finding is consistent with the work of Desai et al. (2010), when the retailers engage a general level of competition with one another. The rationality is that the wholesale price committed in advance by the manufacturer undercuts the retailers' market power and hence distorts each retailer's best response for a given time horizon, in that commitment contracts eliminate strategic inventories. Only the dynamic contract will trigger the retailer's strategic inventory that restores the fare loss caused by the committed wholesale prices.

To summarize this subsection, for each individual supply chain party to achieve a higher profit by invoking retailers' strategic inventories, both the manufacturer and retailers must collectively comply with the necessary conditions established by Proposition 1. Furthermore, Proposition 1 in part, also underscores how the manufacturer as a dominant player is capable of exploiting his counterparts, i.e., the retailers, because of his strictly dominant market power. In other words, the manufacturer's choice of contract could single-handedly dictate the retailers' payoff.

However, it is imperative to recognize that the manufacturer's ability to eliminate the effective inventory strategy of his counterparts does not necessarily suggest the manufacturer's motivation to do so. Rather, the outcome of such a course of action would result in both parties and the entire system being worse off, according to Proposition 2.

It is also crucial to note that the previous study conducted by Anand et al. (2008) has demonstrated that, in a vertical supply chain as a whole of a singular manufacturer and a singular retailer, both firms individually and the system will do strictly better with a dynamic contract than with a commitment contract, if holding inventory is not too costly. Our findings show strong consistency in terms of the dynamic contract and strategic inventory's economic and social advantages. Therefore, in greater generality, we conclude that the dynamic contract and strategic inventory Pareto-dominates the commitment contract for all supply chain parties, the channel, and society.

### 3.2 Comparison of Downstream Monopoly vs. Duopoly

In this subsection, we focus on the performances of each supply chain entity and the whole system for monopolistic and duopolistic retailer cases. Our results from the retailer duopoly are analogous to those of retailer monopoly by Anand et al. (2008).

Table 8 first compares the results of a supply chain with a monopolistic retailer against a pair of duopolistic retailers when the inventory holding cost is sufficiently low, i.e., $h<a/12$, in which case, strategic inventory will be triggered according to Proposition 1. First, we find the downstream duopoly suppresses the double marginalization effect, i.e. ${}^{D}p_{avg}^{Duo} < {}^{D}p_{avg}^{Mon}$. Second, noticeably we find the manufacturer's total profit is increased by the duopolistic retailers, i.e. ${}^{D}\pi M_{avg}^{Duo} > {}^{D}\pi M_{avg}^{Mon}$. The most pronounced result is the transfer of retailer's profit to the manufacturer, as we notice ${}^{D}\pi R_{total}^{Duo} < 2\,{}^{D}\pi R_{total}^{Duo} < {}^{D}\pi R_{total}^{Mon}$; that is, when an incumbent monopolistic retailer is joined by her peer to form a duopoly, not only would her profit be drastically shrinking, but also the total downstream profit of the duopolists.

Why is the downstream competition culpable for the retailers? The rationale can be three-fold. Consider first the Cournot competition between the two retailers. The Cournot model stipulates that each competitor's only strategy is to choose her outputs during the selling season. Consequently, neither retailer could cut the output in order to capture more consumer surplus, as doing so will benefit her counterpart while marginalizing her own gain.

Consider next each retailer's market-power shift from the monopoly to the duopoly formed. Because of the presence of one additional player in the system, a shrinkage of the incumbent retailer's market power is the only plausible result, which will be further illustrated in the next subsection. In particular, to counteract the manufacturer, a duopolistic retailer has a narrower range of output choice, since the output cut that would have been effective for her in monopoly is now reprehensive. Hence, as each duopolist's market power has weakened, the dynamic contract becomes too powerful for retailers to counteract. In terms of a retailer's profit, the formation of a duopoly diminishes the benefit of the strategic inventory.

Lastly, we shall not neglect the retailer's alternative strategy. In other words, could any retailer in duopoly have done differently to ameliorate her profit? The answer is no, in that the subgame-perfect equilibrium, in this case, is also Pareto-dominating; any deviated strategy for the retailer, i.e., carrying no inventory would be rendered worse off. (See Table 3 and Table 4.)

Further observations from Table 8 indicate that despite the shrinking profit of the duopolistic retailers compared to that of the monopolist, the channel profit can still benefit from the downstream competition, since ${}^{D}\pi_{total}^{Duo} > {}^{D}\pi_{total}^{Mon}$. This result further suggests that the manufacturer is the major gainer from such competition, and that such gain can be viewed as retailers' loss of surplus transferred from downstream to upstream. Furthermore, we find that social welfare is also improved by the downstream competition, as ${}^{D}SW_{total}^{Duo} > {}^{D}SW_{total}^{Mon}$.

Table 9 next compares the result of a supply chain with a retailer monopoly and duopoly when the inventory holding cost is high, i.e., a/12<h<a/4. Notice, even within different h range, the measures in Table 9 are identical to those in Table 8 for the monopoly model. Table 10 compares the outcomes from a monopolistic retailer against a pair of duopolistic retailers under a commitment contract. Such a strategy eliminates the downstream retailer's inventory according to Proposition 1. The comparisons in Table 9 and Table 10 indicate a strong similarity to our findings from Table 8. Hence, we conclude in broader generality that the downstream retailer's competition (under both contract types) reduces the average product price by weakening the double-marginalization effect, improves the manufacturer's profit, undermines the retailers' profit, and benefits the entire supply chain's profit and the social welfare.

### 3.3 Retailer's Market Power

In this subsection, we discuss how carrying strategic inventory affects each player's market power in the supply chain model presented. A key measure to a firm's market power in a static market is the Lerner Index (Lerner, 1934), defined as the margin between the price and the marginal cost, i.e., L=(P-MC)/P, where P is the price of a product and MC is its marginal cost.

In this paper, all Lerner Indices are computed by using the weighted average of retail and wholesale prices from each given setting. It is straightforward to show that the manufacturer's monopoly power is 1, a constant that is independent of strategic inventory carried by the retailers, indicating the manufacturer's unwavering dominance when downstream retailers exercise inventory strategies. The more delicate result is the market power change of each retailer due to the change of inventory holding cost, h, shown in Table 2. Particularly, when $h \geq \frac{a}{12}$ the indices collapse into those of the commitment contract, according to Lemma 2.

*Table 2 A Retailer's Lerner Index (h≤a)*

| | Dynamic (D) | Commitment (C) |
|---|---|---|
| **Duopoly (Duo)** | $\frac{4073a^2 - 1032ah - 20592h^2}{2(7849a^2 - 1632ah - 3600h^2)}$ | $\frac{1}{4}$ |
| **Monopoly (Mon)** | $\frac{155a^2 + 52ah - 376h^2}{461a^2 - 84ah - 104h^2}$ | $\frac{1}{3}$ |

First, we compare the market power of a single retailer of duopoly against that of monopoly (Anand et al., 2008) under each wholesale contract. For simplicity, we limit the holding cost within 0<h<a/12 in the dynamic contract situation, despite that a monopolistic retailer's strategic inventory will still be triggered when a/12<h<4/a. Not surprisingly, we find each retailer's market power diminishes as the number of identical retailers increases, i.e., ${}^{D}L^{Mon} > {}^{D}L^{Duo}$ and ${}^{C}L^{Mon} > {}^{C}L^{Duo}$, notwithstanding the upstream manufacturer's choice of contract.

Next, we compare each retailer's market power under a dynamic contract against commitment contract in each type of downstream competition. Recall Proposition 1, to invoke strategic inventory, the retailers must maintain the holding cost sufficiently low, i.e., 0<h<a/12. Since ${}^{D}L^{Duo} > {}^{C}L^{Duo}$, and ${}^{D}L^{Mon} > {}^{C}L^{Mon}$, we conclude that the dynamic contract and its resultant strategic inventory strengthen each retailer's market power.

Furthermore, we observe that each retailer has a different inventory holding cost threshold in monopolistic and duopolistic downstream competition; any holding cost of h≥a/4 would eliminate strategic inventory for the monopolistic retailer, whereas any holding cost h≥a/12 would do so for a duopolistic retailer, as was also established by Lemma 1. Hence, we can rewrite the Lerner Index for each competition under the dynamic contract in the following function with respect to the holding cost ratio, h/a.

$$^{D}L^{Mon}(x) = \begin{cases} \frac{155+52x-376x^2}{461-84x-104x^2}, 0 < x < \frac{1}{4} \\ \frac{1}{3} \qquad\qquad\qquad\qquad , x \geq \frac{1}{4} \end{cases}$$

$$^{D}L^{Duo}(x) = \begin{cases} \frac{4073-1032x-20592x^2}{2(7849-1632x-3600x^2)}, 0 < x < \frac{1}{12} \\ \frac{1}{4} \qquad\qquad\qquad\qquad , x \geq \frac{1}{12} \end{cases}$$

$$\text{Where } x = \frac{h}{a}$$

For both monopoly and duopoly, the Lerner Index of a retailer at equilibrium reaches the minima and sustains such levels under one of the following two conditions: (a) when holding inventory is too costly, or (b) when the upstream manufacturer adopts a commitment wholesale contract, as is evident in Proposition 1. We graph the Lerner Indices in both cases in Figure 2.

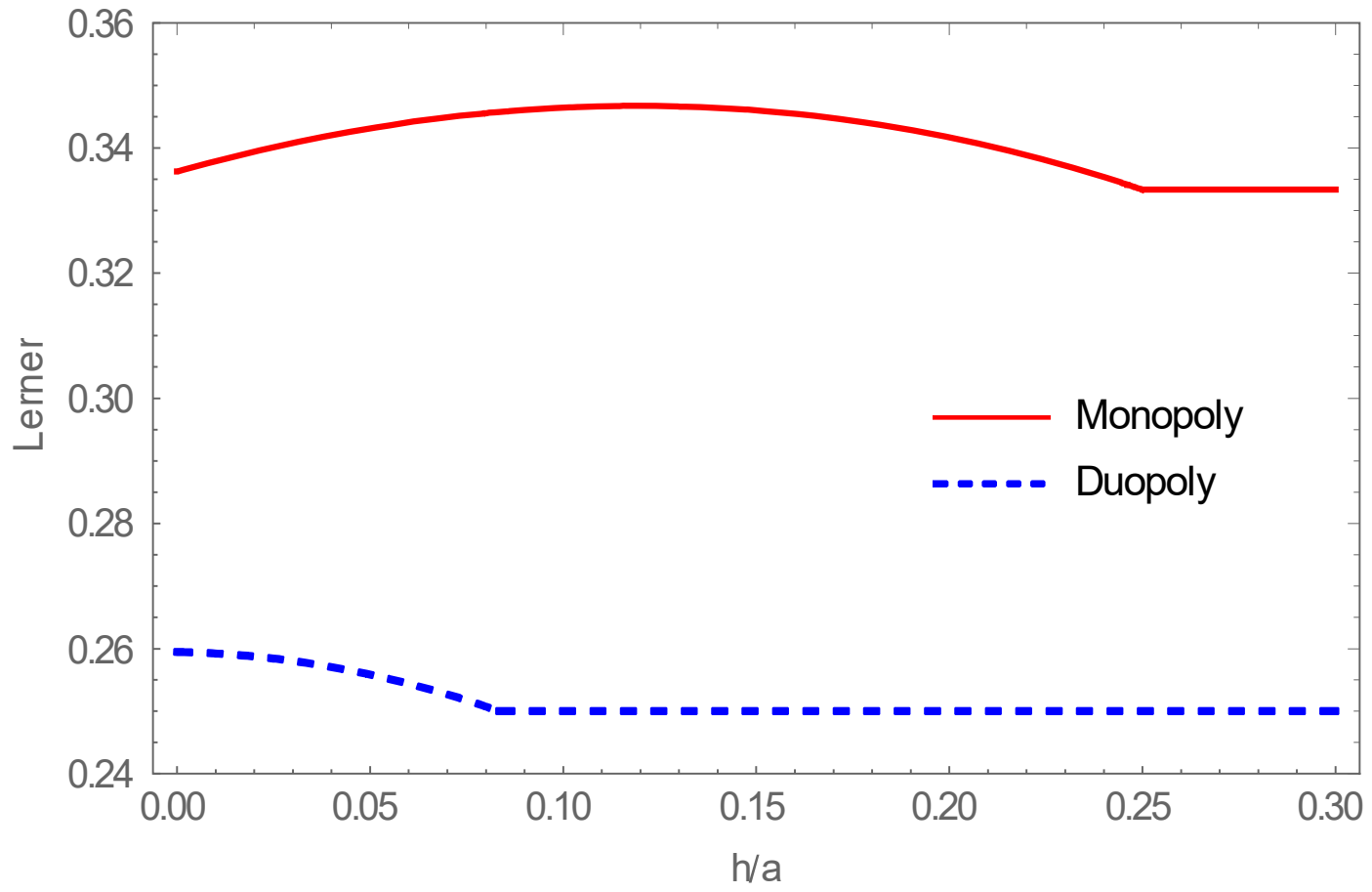

*Figure 2 A Retailer's Market Power in Supply Chain*

### 3.4 Inventory Holding Cost

In the previous subsection, we have compared and demonstrated the superiority of dynamic contract and strategic inventory in bolstering the supply chain's welfare. In this subsection, we will discuss the impact of inventory holding cost rate, h, to both theoretical and practical extent.

Both Proposition 1 and Anand et al. (2008) indicate that the inventory holding cost rate is important to the system's SPNE. Although Anand and Arya et al., (2013) consider the value of h in the assumed range of 0<h<a/4, little literature so far has proffered analytical insights for the choice of this holding cost range. Our further exploration of Anand's work shows that when such an assumption is relaxed, a similar pattern of equilibrium solution resembling Lemma 1 emerges in the retailer's monopoly model. (See Appendix C and D.) When $h \geq a/4$, the SPNE under dynamic contracts becomes identical to what is under commitment contracts. In other words, the elimination of the inventory becomes a self-enforcing agreement. We conjecture that in any given vertical supply chain model with one manufacturer and any number of retailers, there exists a critical value, $c_0$, a *Costliness Threshold*, such that when $h/a \geq c_0$, a downstream retailer carries no inventory. The ratio $h/a$, which is also frequently referred to in this study, can be interpreted as the *Costliness Threshold* of holding inventories.

In subsection 3.3, we have shown that the costliness threshold dictates the market power of a retailer, as is depicted in Figure 2. In a retailer monopoly setting, $c_0$ holds at $1/4$, whereas in a retailer duopoly setting, $c_0$ holds at $1/12$. Furthermore, similar to Table 1 in subsection 3.2, Table 7 in Appendix E can also be reproduced by setting up the problem with a dynamic contract case under different levels of costliness threshold (i.e., h/a<12 vs. h/a≥12). Clearly, the costliness threshold of holding inventories is a key variable that reins in the equilibrium solution.

In practice, when it comes to supply chain planning, businesses merely rely on heuristics to determine appropriate inventory holding cost range. The norm often links the holding cost to the value of inventory with a ratio. Anecdotal evidence suggests that a 25% of inventory value can be used in most industries as a rule-of-thumb (REM Associates). To evaluate if Anand's (2008) assumption can be coarsely validated by such evidence, we simply need to examine the equilibrium solutions. In the supply chain models, the wholesale price $w_{avg}$ appropriately represents the inventory value. Clearly, in JIT cases, $w_{avg} = a/2$ holds true in both retailer monopoly and duopoly. Thus, $25\%w_{avg}$ yields $a/8$, falling well within the range of 0<h<a/4. In non-JIT (dynamic contract) case, because the value of $w_{avg}$ and range of holding cost is codependent, gaining insights with respect to the wholesale price and holding cost from these quantities is difficult. Nonetheless, one could still observe that $w_1 > \frac{a}{2} > w_2$ holds with h<a/12 in a duopoly, and h<a/4 in a monopoly.

## 4. Conclusion

The principal objective of this study is to examine how the retailer's behavior in utilizing inventory as a strategy is affected by and impacts all other parties in a supply chain. Our analysis relies on a game-theoretic model, in which a supply chain consists of a singular upstream manufacturer and a pair of identical downstream retailers. We also broaden the scope of the retailer monopoly framework constructed by previous researchers by expanding the retailer's inventory holding cost range and the downstream competition type.

Overall, two perspectives emerge regarding the strategic inventory in our retailer duopoly framework. The first perspective concerns the effect of strategic inventories and the conditions under which such inventories exist. We find that the dynamic contract and strategic inventory remains an effective course of action for not only the retailers, but also the manufacturer, consumers, the channel, and the whole society since it Pareto-dominates the commitment contract and JIT operation. However, the existence of strategic inventory requires a pair of necessary-and-sufficient conditions: a relatively low costliness of holding inventory by the retailers, and a dynamic wholesale contract employed by the upstream manufacturer.

The second perspective concerns how the type of downstream competition affects each firm in the supply chain. Our most significant finding in such perspective is that the Cournot-like duopoly between the downstream retailers will benefit each individual supply chain party and society as a whole but the retailers themselves, when strategic inventories are carried. In essence, the downstream competition transfers the profits of retailers to the upstream manufacturer.

## Appendices

### A. Proof of Lemma 1 (Dynamic Contract, Duopoly Retailers)

In all cases, we assume all of the following conditions holds true.

$0< w_1< a$, $0< w_2< a$, $I_1, I_2\geq 0$, $h>0$, $a-b(I_1+I_2)\geq 0$.

**Case (1)** Domain Conditions: $(\frac{a-w_2}{3b}-I_1<0, \frac{a-w_2}{3b}-I_2<0)$

In period 2, retailer 1 and 2's profits are:

$$\Pi R_{21}=[a-b(q_{21}+q_{22}+I_1+I_2)](q_{21}+I_1)-w_2q_{21};$$

$$\Pi R_{22}=[a-b(q_{21}+q_{22}+I_1+I_2)](q_{22}+I_2)-w_2q_{22}.$$

The FOC and SOC of each retailer are as follows, respectively.

$$\frac{\partial \Pi R_{21}}{\partial q_{21}}=a-b(I_1+q_{21})-b(I_1+I_2+q_{21}+q_{22})-w_2, \frac{\partial^2 \Pi R_{21}}{\partial q_{21}^2}=-2b<0;$$

$$\frac{\partial \Pi R_{22}}{\partial q_{22}}=a-b(I_2+q_{22})-b(I_1+I_2+q_{21}+q_{22})-w_2, \frac{\partial^2 \Pi R_{22}}{\partial q_{22}^2}=-2b<0.$$

Let $\frac{\partial \Pi R_{21}}{\partial q_{21}}=0$, and $\frac{\partial \Pi R_{22}}{\partial q_{22}}=0$. Solve for $q_{21}$ and $q_{22}$ simultaneously. The solutions are

$$q_{21}^0=\frac{a-w_2}{3b}-I_1, q_{22}^0=\frac{a-w_2}{3b}-I_2$$

Given the domain conditions, $q_{21}^0$ and $q_{22}^0$ are both non-positive. Hence, we denote the 2nd period optimal order quantities of retailers:

$$q_{21}^*=q_{22}^*=0$$

In period 2, the manufacturer's profit is:

$$\Pi M_2=w_2*(q_{21}+q_{22})$$

Given the optimal value of $q_{21}^*$ and $q_{22}^*$, we can have the following but later we will restrict $w_2{}^*$ more precisely:

$$o\leq w_2{}^*\leq a$$

In period 1, each retailer decides the ordering quantity and inventory to carry to the next period. Retailer 1 and 2's profit in current period can be written as

$$\pi R_{11}=[a-b\times(q_{11}+q_{12})]\times q_{11}-[w_1\times(q_{11}+I_1)]-h\times I_1$$

$$\pi R_{12}=[a-b\times(q_{11}+q_{12})]\times q_{12}-[w_1\times(q_{12}+I_2)]-h\times I_2$$

Because each retailer's objective in period 1 is to maximise the profit of both periods (i.e., the entire time horizon), we set up and simultaneously solve equations

$$\frac{\partial(\Pi R_{11}+\Pi R_{21})}{\partial q_{11}}=0 \text{ and } \frac{\partial(\Pi R_{12}+\Pi R_{22})}{\partial q_{12}}=0, \text{ yielding}$$

$$q_{11}^0=q_{12}^0=\frac{a-w_1}{3b}.$$

It is easy to verify that both SOC's are negative. Since $(a-w_1)>0$, $q_{11}^0$and $q_{12}^0$ are both strictly positive. Hence, we denote the 1st period optimal order quantities of retailers:

$$q_{11}^*=q_{12}^*=\frac{a-w_1}{3b}.$$

Similarly, to derive the profit maximising inventory level, we first examine FOC and SOC, noticing

$$\frac{\partial(\Pi R_{11}+\Pi R_{21})}{\partial I_1} = a - h - b\text{I}1 - b(I_1 + I_2) - w_1,$$

$$\frac{\partial(\Pi R_{12}+\Pi R_{22})}{\partial I_2} = a - h - b\text{I}2 - b(I_1 + I_2) - w_1,$$

$$\frac{\partial^2(\Pi R_{11}+\Pi R_{21})}{\partial I_1^2} = -2b, \frac{\partial^2(\Pi R_{12}+\Pi R_{22})}{\partial I_2^2} = -2b.$$

Let $\frac{\partial(\Pi R_{11}+\Pi R_{21})}{\partial I_1} = 0$, and $\frac{\partial(\Pi R_{12}+\Pi R_{22})}{\partial I_2} = 0$. Solve for $I_1$ and $I_2$ simultaneously. The solutions are

$$\text{I}_1^0 = \frac{a-h-w_1}{3b}, I_2^0 = \frac{a-h-w_1}{3b}$$

Given the domain conditions, $\text{I}_1^0$ and $\text{I}_2^0$ are positive and valid. Hence, the optimal solution of $I_1$ and $I_2$ are:

$$\text{I}_1^* = \frac{a-h-w_1}{3b}, I_2^* = \frac{a-h-w_1}{3b}$$

At the beginning of period 1, the manufacturer's decision is the profit-maximising wholesale prices of each period, $\text{w}_1$ and $\text{w}_2$. Each period's profit is

$$\Pi M_1 = w_1(q_{11} + q_{12} + I_1 + I_2) = w_1(-\frac{2(-a + w_1)}{3b} - \frac{2(-a + h + w_1)}{3b})$$

$$\Pi M_2 = w_2(q_{21} + q_{22}) = 0$$

FOC and SOC are respectively

$$\frac{\partial(\Pi M_1+\Pi M_2)}{\partial w_1} = -\frac{4\text{w}1}{3b} - \frac{2(-a+\text{w}1)}{3b} - \frac{2(-a+h+\text{w}1)}{3b}, \frac{\partial(\Pi M_1+\Pi M_2)}{\partial w_2} = 0$$

$$\frac{\partial^2(\Pi M_1+\Pi M_2)}{\partial w_1^2} = -\frac{8}{3b}, \frac{\partial^2(\Pi M_1+\Pi M_2)}{\partial w_2^2} = 0$$

Set up and solve equation

$$\frac{\partial(\Pi M_1+\Pi M_2)}{\partial w_1} = 0$$

We at last obtain $w_1^* = \frac{2a-h}{4}$. $w_2^*$ could take any value as long as $w_2^* \geq \frac{a}{2} + \frac{3h}{4}$.

Inserting the equilibrium solution yields the manufacturer and each retailer's total profits[2]:

$$\Pi M^{(1)} = \frac{(2a - h)^2}{12b}$$

$$\Pi R^{(1)} = \frac{4a^2 - 4ah + 5h^2}{72b}$$

**Case (2)** Domain Conditions: $(\frac{a-w_2}{3b} - I_1 \geq 0, \frac{a-w_2}{3b} - I_2 \geq 0), (w_1^* > \frac{7a}{12} - h) \equiv (h > \frac{a}{12})$

[2] We use the superscription (n) to denote the total profit in case n, e.g., $\Pi R^{(1)}$ denotes the total profit of each retailer in case 1.

Similar to the previous case, we find the following solutions for the retailers in the second period and given the domain conditions, they are optimal values:

$$q_{21}^{*} = \frac{a-w_2}{3b} - I_1,\ q_{22}^{*} = \frac{a-w_2}{3b} - I_2$$

In period 2, the manufacturer's profit is:

$$\Pi M_2 = w_2 * (q_{21} + q_{22})$$

FOC and SOC are respectively:

$$\frac{\partial(\Pi M_2)}{\partial w_2} = -\frac{2w2}{3b} - \frac{-a + 3bI1 + w2}{3b} - \frac{-a + 3bI2 + w2}{3b}$$

$$\frac{\partial^2(\Pi M_2)}{\partial w_2^2} = -\frac{4}{3b}$$

Set up and solve equation:

$$\frac{\partial(\Pi M_2)}{\partial w_2} = 0$$

We obtain the following optimal solution, which is non-negative given the domain condition:

$$W_2^{*} = \frac{1}{4}(2a - 3bI_1 - 3bI_2)$$

In period 1, each retailer decides the ordering quantity and inventory to carry to the next period. Retailer 1 and 2's profit in current period can be written as

$$\pi R11 = (a - b * (q11 + q12)) * q11 - (w1 * (q11 + I1)) - hI1$$

$$\pi R12 = (a - b * (q11 + q12)) * q12 - (w1 * (q12 + I2)) - hI2$$

Because each retailer's objective in period 1 is to maximise the profit of both periods (i.e., the entire time horizon), we set up and simultaneously solve equations

$\frac{\partial(\Pi R_{11}+\Pi R_{21})}{\partial q_{11}} = 0$ and $\frac{\partial(\Pi R_{12}+\Pi R_{22})}{\partial q_{12}} = 0$, yielding

$$q_{11}^{*} = q_{12}^{*} = \frac{a-w_1}{3b}.$$

Similarly, to derive the profit maximising inventory level, we first examine FOC and SOC such that,

$$\frac{\partial(\Pi R_{11}+\Pi R_{21})}{\partial I_1} = 0,$$

$$\frac{\partial(\Pi R_{12}+\Pi R_{22})}{\partial I_2} = 0,$$

$$\frac{\partial^2(\Pi R_{11}+\Pi R_{21})}{\partial I_1^2} = -2b, \frac{\partial^2(\Pi R_{12}+\Pi R_{22})}{\partial I_2^2} = -2b.$$

By Solving FOC's simultaneously for $I_1$ and $I_2$ the following are obtained:

$$I_1^0 = \frac{7a - 12h - 12w1}{24b}, I_2^0 = \frac{7a - 12h - 12w1}{24b}$$

Given the domain conditions, $I_1^0$ and $I_2^0$ would be negative. Hence the optimal value of Strategic inventory of retailers are:

$$I_1^* = 0, I_2^* = 0 => W_2^* = \frac{a}{2}$$

At the beginning of period 1, the manufacturer's decision is the profit-maximising wholesale prices of each period, $w_1$ and $w_2$. Each period's profit is

$$\Pi M_1 = w_1(q_{11} + q_{12} + I_1 + I_2) = \frac{2(a - w1)w1}{3b}$$

$$\pi M2 = w2(q21 + q22) = \frac{a^2}{6b}$$

FOC and SOC are respectively

$$\frac{\partial(\Pi M_1 + \Pi M_2)}{\partial w_1} = \frac{2(a - w1)}{3b} - \frac{2w1}{3b}$$

$$\frac{\partial^2(\Pi M_1 + \Pi M_2)}{\partial w_1^2} = -\frac{4}{3b}$$

Solving the FOC gives the following optimal value for W1:

$$w_1^* = \frac{a}{2}$$

Inserting the equilibrium solution yields the manufacturer and each retailer's total profits:

$$\Pi M^{(2)} = \frac{a^2}{3b}$$

$$\Pi R^{(2)} = \frac{a^2}{18b}$$

**Case (3)** Domain Conditions: $(\frac{a-w_2}{3b} - I_1 \geq 0, \frac{a-w_2}{3b} - I_2 \geq 0), (w_1^* \leq \frac{7a}{12} - h) \equiv (h \leq \frac{a}{12})$

All results in this case hold the same up to the optimal value of strategic inventory. Given the domain conditions, we obtain the following for the optimal value of $I_1$ and $I_2$:

$$I_1^* = \frac{7a - 12h - 12w1}{24b}, I_2^* = \frac{7a - 12h - 12w1}{24b}$$

At the beginning of period 1, the manufacturer's decision is the profit-maximising wholesale prices of each period, $w_1$ and $w_2$. Each period's profit is

$$\pi M1 = w1 * (q11 + q12 + I1 + I2) = (-\frac{-7a + 3(2a - h) + 12h}{12b} + \frac{2(a + \frac{1}{4}(-2a + h))}{3b})w1$$

$$\pi M2 = w2(q21 + q22) =$$

$$-\frac{(2a + \frac{1}{4}(-7a + 3(2a - h) + 12h))(-a + \frac{1}{8}(7a - 3(2a - h) - 12h) + \frac{1}{4}(2a + \frac{1}{4}(-7a + 3(2a - h) + 12h)))}{6b}$$

FOC and SOC of the total profit are respectively

$$\frac{\partial(\Pi M_1 + \Pi M_2)}{\partial w_1} = \frac{49a^2 + 126ah + 81h^2 + 160aw1 - 224hw1}{384b}$$

$$\frac{\partial^2(\Pi M_1 + \Pi M_2)}{\partial w_1^2} = -\frac{4}{3b}$$

Solving the FOC gives the following optimal value for W1:

$$W_1^* = \frac{3}{124}(21a - 4h)$$

Inserting the equilibrium solution yields the manufacturer and each retailer's total profits:

$$\Pi M^{(3)} = \frac{125a^2 - 24ah + 144h^2}{372b}$$

$$\Pi R^{(3)} = \frac{4073a^2 - 3636ah + 10656h^2}{69192b}$$

*Table 3 Equilibria and profits of the supply chain with retailer duopoly under dynamic contract*

| Case | h | Solution | Profits |
|---|---|---|---|
| (1) | $h < \frac{2}{3}a$ | $\{w_1^*, w_2^*\} = \{\frac{2a-h}{4}, w_2\}(\forall w_2 > \frac{a}{2} + h)$, $\{I_1^*, I_2^*\} = \{\frac{2a-3h}{12b}, \frac{2a-3h}{12b}\}$, $\{q_{11}^*, q_{12}^*, q_{21}^*, q_{22}^*\} = \{\frac{2a+h}{12b}, \frac{2a+h}{12b}, 0,0\}$ | $\Pi M^{(1)} = \frac{(2a-h)^2}{12b}$ $\Pi R^{(1)} = \frac{4a^2 - 4ah + 5h^2}{72b}$ |
| (2) | $h \geq \frac{a}{12}$ | $\{w_1^*, w_2^*\} = \{\frac{a}{2}, \frac{a}{2}\}$, $\{I_1^*, I_2^*\} = \{0,0\}$, $\{q_{11}^*, q_{12}^*, q_{21}^*, q_{22}^*\} = \{\frac{a}{6b}, \frac{a}{6b}, \frac{a}{6b}, \frac{a}{6b}\}$ | $\Pi M^{(2)} = \frac{a^2}{3b}$ $\Pi R^{(2)} = \frac{a^2}{18b}$ |
| (3) | $h < \frac{a}{12}$ | $\{w_1^*, w_2^*\} = \{\frac{3(21a-4h)}{124}, \frac{55a+84h}{124}\}$, $\{I_1^*, I_2^*\} = \{\frac{7(a-12h)}{186b}, \frac{7(a-12h)}{186b}\}$, $\{q_{11}^*, q_{12}^*, q_{21}^*, q_{22}^*\} = \{\frac{61a+12h}{372b}, \frac{61a+12h}{372b}, \frac{55a+84h}{372b}, \frac{55a+84h}{372b}\}$ | $\Pi M^{(3)} = \frac{125a^2 - 24ah + 144h^2}{372b}$ $\Pi R^{(3)} = \frac{4073a^2 - 3636ah + 10656h^2}{69192b}$ |

When the ranges of the holding cost overlap in Table 3, the following can be easily shown, and we have two possible subgame perfect equilibria, cases (2) and (3).

$$\Pi M^{(1)} \leq \Pi M^{(2)}, \ \Pi R^{(1)} \leq \Pi R^{(2)}$$

$$\Pi M^{(1)} \leq \Pi M^{(3)}, \ \Pi R^{(1)} \leq \Pi R^{(3)}$$

**B. Proof of Lemma 2 (Commitment Contract, Duopoly Retailers)**

In all cases, we assume all of the following conditions holds true.

$0< w_1< a$, $0< w_2< a$, $I_1, I_2\geq 0$, $h>0$, $a-b\,(I_1+I_2) \geq 0$.

**Case (1)** Domain Conditions: $\frac{a-w_2}{3b} - I_1 \geq 0, \frac{a-w_2}{3b} - I_2 \geq 0, a - w_1 - h \geq 0$

In period 2, retailer 1 and 2's profits are

$$\Pi R_{21}=[a - b(q_{21} + q_{22} + I_1 + I_2)](q_{21} + I_1) - w_2 q_{21};$$

$$\Pi R_{22}=[a - b(q_{21} + q_{22} + I_1 + I_2)](q_{22} + I_2) - w_2 q_{22}.$$

The FOC and SOC of each retailer are as follows, respectively.

$$\frac{\partial \Pi R_{21}}{\partial q_{21}} = a - b(I_1 + q_{21}) - b(I_1 + I_2 + q_{21} + q_{22}) - w_2, \frac{\partial^2 \Pi R_{21}}{\partial q_{21}^2} = -2b < 0;$$

$$\frac{\partial \Pi R_{22}}{\partial q_{22}} = a - b(I_2 + q_{22}) - b(I_1 + I_2 + q_{21} + q_{22}) - w_2, \frac{\partial^2 \Pi R_{22}}{\partial q_{22}^2} = -2b < 0.$$

Let $\frac{\partial \Pi R_{21}}{\partial q_{21}} = 0$, and $\frac{\partial \Pi R_{22}}{\partial q_{22}} = 0$. Solve for q21 and q22 simultaneously. The solutions are

$$q_{21}^0 = -\frac{-a+3bI1+w2}{3b}, q_{22}^0 = -\frac{-a+3bI2+w2}{3b}$$

Since $\frac{a-w_2}{3b} - I_1 \geq 0, \frac{a-w_2}{3b} - I_2 \geq 0$, $q_{21}^0$ and $q_{22}^0$ are both non-positive. Hence, we denote the 2nd period optimal order quantities of retailers:

$$q_{21}^* = q_{22}^* = 0.$$

In period 1, each retailer decides the ordering quantity and inventory to carry to the next period. Retailer 1 and 2's profit in current period can be written as

$$\pi R_{11} = [a - b * (q_{11} + q_{12})] * q_{11} - [w_1 * (q_{11} + I_1)] - h * I_1$$

$$\pi R_{12} = [a - b * (q_{11} + q_{12})] * q_{12} - [w_1 * (q_{12} + I_2)] - h * I_2$$

Because each retailer's objective in period 1 is to maximise the profit of both periods (i.e., the entire time horizon), we set up and simultaneously solve equations

$$\frac{\partial(\Pi R_{11}+\Pi R_{21})}{\partial q_{11}} = 0 \text{ and } \frac{\partial(\Pi R_{12}+\Pi R_{22})}{\partial q_{12}} = 0, \text{ yielding}$$

$$q_{11}^0 = q_{12}^0 = \frac{a-w_1}{3b}.$$

It is easy to verify that both SOC's are negative. Since $a - w_1 > 0$, $q_{11}^0$and $q_{12}^0$ are both strictly positive. Hence we denote the 1st period optimal order quantities of retailers:

$$q_{11}^* = q_{12}^* = \frac{a-w_1}{3b}.$$

Similarly, to derive the profit maximising inventory level, we first examine FOC and SOC, noticing

$$\frac{\partial(\Pi R_{11}+\Pi R_{21})}{\partial I_1} = a - h - bI_1 - b(I_1 + I_2) - w_1,$$

$$\frac{\partial(\Pi R_{12}+\Pi R_{22})}{\partial I_2} = a - h - bI_2 - b(I_1 + I_2) - w_1,$$

$$\frac{\partial^2(\Pi R_{11}+\Pi R_{21})}{\partial I_1^2} = -2b, \frac{\partial^2(\Pi R_{12}+\Pi R_{22})}{\partial I_2^2} = -2b.$$

Let $\frac{\partial(\Pi R_{11}+\Pi R_{21})}{\partial I_1}=0$, and $\frac{\partial(\Pi R_{12}+\Pi R_{22})}{\partial I_2}=0$. Solve for $I_1$ and $I_2$ simultaneously. The solutions are

$$\mathrm{I}_1^0=\frac{a-h-\mathrm{w}_1}{3b}, I_2^0=\frac{a-h-\mathrm{w}_1}{3b}$$

Because $a-w_1-h\geq 0$, $\mathrm{I}_1^0\geq 0$ and $I_2^0\geq 0$. Hence, the optimal solution of $\mathrm{I}_1$ and $\mathrm{I}_2$ are

$$\mathrm{I}_1^*=\frac{a-h-\mathrm{w}_1}{3b}, I_2^*=\frac{a-h-\mathrm{w}_1}{3b}$$

At the beginning of period 1, the manufacturer's decision is the profit-maximising wholesale prices of each period, $\mathrm{w}_1$ and $\mathrm{w}_2$. Each period's profit is

$$\pi\mathrm{M}_1=\mathrm{w}_1(q_{11}+q_{12}+I_1+I_2)=\mathrm{w1}(-\frac{2(-a+\mathrm{w}_1)}{3b}-\frac{2(-a+h+\mathrm{w}_1)}{3b})$$

$$\pi\mathrm{M}_2=\mathrm{w}_2(q_{21}+q_{22})=0$$

FOC and SOC are respectively

$$\frac{\partial(\Pi M_1+\Pi M_2)}{\partial w_1}=-\frac{4\mathrm{w}_1}{3b}-\frac{2(-a+\mathrm{w}_1)}{3b}-\frac{2(-a+h+\mathrm{w}_1)}{3b}, \frac{\partial(\Pi M_1+\Pi M_2)}{\partial w_2}=0$$

$$\frac{\partial^2(\Pi M_1+\Pi M_2)}{\partial w_1^2}=-\frac{8}{3b}, \frac{\partial^2(\Pi M_1+\Pi M_2)}{\partial w_2^2}=0$$

Set up and solve equation

$$\frac{\partial(\Pi M_1+\Pi M_2)}{\partial w_1}=0$$

We at last obtain $w_1^*=\frac{2a-h}{4}$. $w_2^*$ could take any value as long as $w_2^*\geq\frac{a}{2}+\frac{3h}{4}$.

With the optimal solution $w_1^*, w_2^*, \mathrm{I}_1^*, I_2^*, \mathrm{q}_{11}^*, \mathrm{q}_{12}^*, \mathrm{q}_{21}^*, \mathrm{q}_{22}^*$, we compute the total profit of manufacturer and (each) retailer as

$$\Pi M^{(1)}=\Pi M_1+\Pi M_2=\frac{(2a-h)^2}{12b}$$

$$\Pi R^{(1)}=\Pi R_{11}+\Pi R_{21}=\Pi R_{12}+\Pi R_{22}=\frac{4a^2-4ah+5h^2}{72b}$$

**Case (2)** Domain Conditions: $\frac{a-w_2}{3b}-I_1<0, \frac{a-w_2}{3b}-I_2<0, w_2<w_1+h$

In period 2, retailer 1 and 2's profits are

$$\Pi R_{21}=[a-b(q_{21}+q_{22}+I_1+I_2)](q_{21}+I_1)-w_2q_{21};$$

$$\Pi R_{22}=[a-b(q_{21}+q_{22}+I_1+I_2)](q_{22}+I_2)-w_2q_{22}.$$

The FOC and SOC of each retailer are as follows, respectively.

$$\frac{\partial\Pi R_{21}}{\partial q_{21}}=a-b(I_1+q_{21})-b(I_1+I_2+q_{21}+q_{22})-w_2, \frac{\partial^2\Pi R_{21}}{\partial q_{21}^2}=-2\mathrm{b}<0;$$
$$\frac{\partial\Pi R_{22}}{\partial q_{22}}=a-b(I_2+q_{22})-b(I_1+I_2+q_{21}+q_{22})-w_2, \frac{\partial^2\Pi R_{22}}{\partial q_{22}^2}=-2\mathrm{b}<0.$$

Let $\frac{\partial\Pi R_{21}}{\partial q_{21}}=0$, and $\frac{\partial\Pi R_{22}}{\partial q_{22}}=0$. Solve for q21 and q22 simultaneously. The solutions are

$$q_{21}^{0} = -\frac{-a+3bI_1+w_2}{3b}, q_{22}^{0} = -\frac{-a+3bI_2+w_2}{3b}$$

Since $\frac{a-w_2}{3b} - I_1 < 0, \frac{a-w_2}{3b} - I_2 < 0$, $q_{21}^{0}$ and $q_{22}^{0}$ are both strictly positive. Hence, we denote the 2nd period optimal order quantities of retailers:

$$q_{21}^{*} = -\frac{-a+3bI1+w2}{3b}, q_{22}^{*} = -\frac{-a+3bI2+w2}{3b}$$

In period 1, each retailer decides the ordering quantity and inventory to carry to the next period. Retailer 1 and 2's profit in current period can be written as

$$\pi R_{11} = [a - b * (q_{11} + q_{12})] * q_{11} - [w_1 * (q_{11} + I_1)] - h * I_1$$

$$\pi R_{12} = [a - b * (q_{11} + q_{12})] * q_{12} - [w_1 * (q_{12} + I_2)] - h * I_2$$

Because each retailer's objective in period 1 is to maximise the profit of both periods (i.e., the entire time horizon), we set up and simultaneously solve equations

$$\frac{\partial(\Pi R_{11}+\Pi R_{21})}{\partial q_{11}} = 0 \text{ and } \frac{\partial(\Pi R_{12}+\Pi R_{22})}{\partial q_{12}} = 0, \text{ yielding}$$

$$q_{11}^{0} = q_{12}^{0} = \frac{a-w_1}{3b}.$$

It is easy to verify that both SOC's are negative. Since $a - w_1 > 0$, $q_{11}^{0}$and $q_{12}^{0}$ are both strictly positive. Hence, we denote the 1st period optimal order quantities of retailers:

$$q_{11}^{*} = q_{12}^{*} = \frac{a-w1}{3b}.$$

Similarly, to derive the profit maximising inventory level, we first examine FOC and SOC, noticing

$$\frac{\partial(\Pi R_{11}+\Pi R_{21})}{\partial I_1} = -h - w_1 + w_2, \frac{\partial(\Pi R_{12}+\Pi R_{22})}{\partial I_2} = -h - w_1 + w_2,$$

$$\frac{\partial^2(\Pi R_{11}+\Pi R_{21})}{\partial I_1^2} = 0, \frac{\partial^2(\Pi R_{12}+\Pi R_{22})}{\partial I_2^2} = 0.$$

Because $w_2 < w_1 + h$, both FOC's are negative, indicating total profit is strictly declining with respect to inventory level. Therefore, the optimal inventory levels are

$$I_1^{*} = I_2^{*} = 0$$

At the beginning of period 1, the manufacturer's decision is the profit-maximising wholesale prices of each period, $w_1$ and $w_2$. Each period's profit is

$$\pi M_1 = w_1(q_{11} + q_{12} + I_1 + I_2)$$

$$\pi M_2 = w_2(q_{21} + q_{22})$$

FOC and SOC are respectively

$$\frac{\partial(\Pi M_1+\Pi M_2)}{\partial w_1} = -\frac{2w1}{3b} - \frac{2(-a+w1)}{3b}, \frac{\partial(\Pi M_1+\Pi M_2)}{\partial w_2} = -\frac{2w2}{3b} - \frac{2(-a+w2)}{3b}$$

$$\frac{\partial^2(\Pi M_1+\Pi M_2)}{\partial w_1^2} = -\frac{4}{3b}, \frac{\partial^2(\Pi M_1+\Pi M_2)}{\partial w_2^2} = -\frac{4}{3b}$$

Set up and simultaneously solve equations

$$\frac{\partial(\Pi M_1+\Pi M_2)}{\partial w_1} = 0 \text{ and } \frac{\partial(\Pi M_1+\Pi M_2)}{\partial w_2} = 0$$

We at last obtain $w_1^* = w_2^* = \frac{a}{2}$.

With the optimal solution $w_1^*, w_2^*, I_1^*, I_2^*, q_{11}^*, q_{12}^*, q_{21}^*, q_{22}^*$, we compute the total profit of manufacturer and (each) retailer as

$$\Pi M^{(2)} = \Pi M_1 + \Pi M_2 = \frac{a^2}{3b}$$

$$\Pi R^{(2)} = \Pi R_{11} + \Pi R_{21} = \Pi R_{12} + \Pi R_{22} = \frac{a^2}{18b}$$

*Table 4 Equilibria and profits of the supply chain with retailer duopoly under commitment contract*

| Case | h | Solution | Profit |
|---|---|---|---|
| (1) | Any h | $\{w_1^*, w_2^*\} = \{\frac{2a-h}{4}, w_2\}(\forall w_2 \geq \frac{a}{2} + \frac{3h}{4})$, $\{I_1^*, I_2^*\} = \{\frac{2a-3h}{12b}, \frac{2a-3h}{12b}\}$, $\{q_{11}^*, q_{12}^*, q_{21}^*, q_{22}^*\} = \{\frac{2a+h}{12b}, \frac{2a+h}{12b}, 0,0\}$ | $\Pi M^{(1)} = \frac{(2a-h)^2}{12b}$ $\Pi R^{(1)} = \frac{4a^2 - 4ah + 5h^2}{72b}$ |
| (2) | Any h | $\{w_1^*, w_2^*\} = \{\frac{a}{2}, \frac{a}{2}\}$, $\{I_1^*, I_2^*\} = \{0,0\}$, $\{q_{11}^*, q_{12}^*, q_{21}^*, q_{22}^*\} = \{\frac{a}{6b}, \frac{a}{6b}, \frac{a}{6b}, \frac{a}{6b}\}$ | $\Pi M^{(2)} = \frac{a^2}{3b}$ $\Pi R^{(2)} = \frac{a^2}{18b}$ |

It can be shown that for any h<a,

$$\Pi M^{(1)} \leq \Pi M^{(2)};$$

$$\Pi R^{(1)} \leq \Pi R^{(2)}.$$

Hence, the solution in case (2) is the unique SPNE.

## C. Results: Dynamic Contract, Monopoly Retailers

*Table 5 Equilibria and profits of the supply chain with retailer monopoly under dynamic contract*

| Case | h | Solution | Profits |
|---|---|---|---|
| (1) | $h \le \frac{2}{3}a$ | $\{w_1^*, w_2^*\} = \{\frac{2a-h}{4}, w_2\}(\forall w_2 \ge \frac{a}{2} + \frac{3h}{4})$, $\{I^*\} = \{\frac{2a-3h}{8b}\}$, $\{q_1^*, q_2^*\} = \{\frac{2a+h}{8b}, 0\}$ | $\Pi M^{(1)} = \frac{4a^2 - 4ah + h^2 - (4a-6h)W_2}{16b}$ $\Pi R^{(1)} = \frac{-4a^2 + 4ah + 19h^2 + (16a-24h)W_2}{64b}$ |
| (2) | $h \ge \frac{a}{4}$ | $\{w_1^*, w_2^*\} = \left\{\frac{a}{2}, \frac{a}{2}\right\}$, $\{I^*\} = \{0\}$, $\{q_1^*, q_2^*\} = \{\frac{a}{4b}, \frac{a}{4b}\}$ | $\Pi M^{(2)} = \frac{a^2}{4b}$ $\Pi R^{(2)} = \frac{a^2}{8b}$ |
| (3) | $h < \frac{a}{4}$ | $\{w_1^*, w_2^*\} = \{\frac{9a-2h}{17}, \frac{6a+10h}{17}\}$ $\{I^*\} = \{\frac{5a-20h}{34b}\}$, $\{q_1^*, q_2^*\} = \{\frac{4a+h}{17b}, \frac{3a+5h}{17b}\}$ | $\Pi M^{(3)} = \frac{9a^2 - 4ah + 8h^2}{34b}$ $\Pi R^{(3)} = \frac{155a^2 - 118h + 304h^2}{1156b}$ |

It is straightforward to show that for any h<a,

$$\Pi M^{(1)} \le \Pi M^{(2)} \le \Pi M^{(3)};$$

## D. Results: Commitment Contract, Monopoly Retailers

*Table 6 Equilibria and profits of the supply chain with retailer monopoly under commitment contract*

| Case | h | Solution | Profits |
|---|---|---|---|
| (1) | $h < \frac{a}{2}$ | $\{w_1^*, w_2^*\} = \left\{\frac{a}{2}, w_2\right\}, (\forall w_2 \geq \frac{a}{2} + h)$<br>$\{\mathrm{I}^*\} = \{\frac{a-2h}{4b}\}$,<br>$\{q_1^*, q_2^*\} = \{\frac{a}{4b}, 0\}$ | $\Pi M^{(1)} = \frac{a^2 - ah}{4b}$<br>$\Pi R^{(1)} = \frac{a^2 - 2ah + h^2}{8b}$ |
| (2) | Any h | $\{w_1^*, w_2^*\} = \left\{\frac{a}{2}, \frac{a}{2}\right\}$,<br>$\{\mathrm{I}^*\} = \{0\}$,<br>$\{q_1^*, q_2^*\} = \{\frac{a}{4b}, \frac{a}{4b}\}$ | $\Pi M^{(2)} = \frac{a^2}{4b}$<br>$\Pi R^{(2)} = \frac{a^2}{8b}$ |

It is straightforward to show that for any h<a,

$$\Pi M^{(1)} \leq \Pi M^{(2)};$$

$$\Pi R^{(1)} \leq \Pi R^{(2)};$$

## E. Full Comparison of Scenarios in Tables

*Table 7 No Inventory (JIT) vs. Inventory Carried in a retailer duopoly*

| | | No Inventory | | Inventory |
|---|---|---|---|---|
| Wholesale Price | $w_1$ | $\frac{a}{2}$ | < | $\frac{3}{124}(21a-4h)$ |
| | $w_2$ | $\frac{a}{2}$ | > | $\frac{1}{124}(55a+84h)$ |
| | $w_{avg}$ | $\frac{a}{2}$ | > | $\frac{125a^2-24ah+144h^2}{4(65a-36h)}$ |
| Retail Price | $p_1$ | $\frac{2a}{3}$ | < | $\frac{1}{186}(125a-12h)$ |
| | $p_2$ | $\frac{2a}{3}$ | > | $\frac{1}{62}(39a+28h)$ |
| | $p_{avg}$ | $\frac{2a}{3}$ | > | $\frac{7849a^2-1632ah-3600h^2}{186(65a-36h)}$ |
| Consumer Purchase Qty | $Q_1$ | $\frac{a}{3b}$ | > | $\frac{61a+12h}{186b}$ |
| | $Q_2$ | $\frac{a}{3b}$ | < | $\frac{23a-28h}{62b}$ |
| | $Q_{total}$ | $\frac{2a}{3b}$ | < | $\frac{65a-36h}{93b}$ |
| Inventory | I | 0 | < | $\frac{7(a-12h)}{186b}$ |
| Retailer (ea.) Sales | $q_1$ | $\frac{a}{6b}$ | > | $\frac{61a+12h}{372b}$ |
| | $q_2+I$ | $\frac{a}{6b}$ | < | $\frac{23a-28h}{124b}$ |
| | $q_{total}$ | $\frac{a}{3b}$ | < | $\frac{65a-36h}{186b}$ |
| Retailer (ea.) Profit | $\pi R_1$ | $\frac{a^2}{36b}$ | > | $\frac{1075a^2+28512ah+56592h^2}{138384b}$ |
| | $\pi R_2$ | $\frac{a^2}{36b}$ | < | $\frac{2357a^2-11928ah-11760h^2}{46128b}$ |
| | $\pi R_{total}$ | $\frac{a^2}{18b}$ | < | $\frac{4073a^2-3636ah+10656h^2}{69192b}$ |
| Manufacturer Profit | $\pi M_1$ | $\frac{a^2}{6b}$ | < | $\frac{3(25a-52h)(21a-4h)}{7688b}$ |
| | $\pi M_2$ | $\frac{a^2}{6b}$ | > | $\frac{(55a+84h)^2}{23064b}$ |
| | $\pi M_{total}$ | $\frac{a^2}{3b}$ | < | $\frac{125a^2-24ah+144h^2}{372b}$ |
| SC Profit | $\pi_1$ | $\frac{2a^2}{9b}$ | > | $\frac{7625a^2-1836ah+31104h^2}{34596b}$ |
| | $\pi_2$ | $\frac{2a^2}{9b}$ | < | $\frac{897a^2-448ah-784h^2}{3844b}$ |
| | $\pi_{total}$ | $\frac{4a^2}{9b}$ | < | $\frac{7849a^2-2934ah+12024h^2}{17298b}$ |
| Consumer Surplus | $CS_1$ | $\frac{a^2}{18b}$ | > | $\frac{(61a + 12h)^2}{69192b}$ |
| | $CS_2$ | $\frac{a^2}{18b}$ | < | $\frac{(23a - 28h)^2}{7688b}$ |
| | $CS_{total}$ | $\frac{a^2}{9b}$ | < | $\frac{4241a^2 - 5064ah + 3600h^2}{34596b}$ |
| Social Welfare | $SW_1$ | $\frac{5a^2}{18b}$ | > | $\frac{18971a^2 - 2208ah + 62352h^2}{69192b}$ |
| | $SW_2$ | $\frac{5a^2}{18b}$ | < | $\frac{2323a^2 - 2184ah - 784h^2}{7688b}$ |
| | $SW_{total}$ | $\frac{5a^2}{9b}$ | < | $\frac{19939a^2 - 10932ah + 27648h^2}{34596b}$ |

*Table 8 Retailer's monopoly vs. duopoly under dynamic contract when holding cost is low (h < a/12)*

| | | Monopoly (Anand 2008) | | Duopoly |
|---|---|---|---|---|
| Wholesale Price | $w_1$ | $\frac{1}{17}(9a-2h)$ | > | $\frac{3}{124}(21a-4h)$ |
| | $w_2$ | $\frac{2}{17}(3a+5h)$ | < | $\frac{1}{124}(55a+84h)$ |
| | $w_{avg}$ | $\frac{9a^2-4ah+8h^2}{19a-8h}$ | < | $\frac{125a^2-24ah+144h^2}{4(65a-36h)}$ |
| Retail Price | $p_1$ | $\frac{1}{17}(13a-h)$ | > | $\frac{1}{186}(125a-12h)$ |
| | $p_2$ | $\frac{1}{34}(23a+10h)$ | > | $\frac{1}{62}(39a+28h)$ |
| | $p_{avg}$ | $\frac{461a^2-84ah-104h^2}{34(19a-8h)}$ | > | $\frac{7849a^2-1632ah-3600h^2}{186(65a-36h)}$ |
| Consumer Purchase Qty | $Q_1$ | $\frac{4a+h}{17b}$ | < | $\frac{61a+12h}{186b}$ |
| | $Q_2$ | $\frac{11a-10h}{34b}$ | < | $\frac{23a-28h}{62b}$ |
| | $Q_{total}$ | $\frac{19a-8h}{34b}$ | < | $\frac{65a-36h}{93b}$ |
| Inventory | I | $\frac{5(a-4h)}{34b}$ | > | $\frac{7(a-12h)}{186b}$ |
| Retailer (ea.) Sales | $q_1$ | $\frac{13a-18h}{34b}$ | > | $\frac{61a+12h}{372b}$ |
| | $q_2+I$ | $\frac{3a+5h}{17b}$ | > (mostly[3]) | $\frac{23a-28h}{124b}$ |
| | $q_{total}$ | $\frac{19a-8h}{34b}$ | > | $\frac{65a-36h}{186b}$ |
| Retailer (ea.) Profit | $\pi R_1$ | $\frac{-13a^2+121ah+302h^2}{578b}$ | < | $\frac{1075a^2+28512ah+56592h^2}{138384b}$ |
| | $\pi R_2$ | $\frac{181a^2-360ah-300h^2}{1156b}$ | > | $\frac{2357a^2-11928ah-11760h^2}{46128b}$ |
| | $\pi R_{total}$ | $\frac{155a^2-118ah+304h^2}{1156b}$ | > | $\frac{4073a^2-3636ah+10656h^2}{69192b}$ |
| Manufacturer Profit | $\pi M_1$ | $\frac{(13a-18h)(9a-2h)}{578b}$ | < (mostly) | $\frac{3(25a-52h)(21a-4h)}{7688b}$ |
| | $\pi M_2$ | $\frac{2(3a+5h)^2}{289b}$ | < | $\frac{(55a+84h)^2}{23064b}$ |
| | $\pi M_{total}$ | $\frac{9a^2-4ah+8h^2}{34b}$ | < | $\frac{125a^2-24ah+144h^2}{372b}$ |
| SC Profit | $\pi_1$ | $\frac{104a^2-67ah+338h^2}{578b}$ | < | $\frac{7625a^2-1836ah+31104h^2}{34596b}$ |
| | $\pi_2$ | $\frac{(11a-10h)(23a+10h)}{1156b}$ | < | $\frac{897a^2-448ah-784h^2}{3844b}$ |
| | $\pi_{total}$ | $\frac{461a^2-254ah+576h^2}{1156b}$ | < | $\frac{7849a^2-2934ah+12024h^2}{17298b}$ |

[3] The inequality holds for the "most" part of the given range, unless h is close to 0.

| | | | | |
|---|---|---|---|---|
| Consumer Surplus | $CS_1$ | $\frac{(4a+h)^2}{578b}$ | < | $\frac{(61a+12h)^2}{69192b}$ |
| | $CS_2$ | $\frac{(11a-10h)^2}{2312b}$ | < | $\frac{(23a-28h)^2}{7688b}$ |
| | $CS_{total}$ | $\frac{185a^2-188ah+104h^2}{2312b}$ | < | $\frac{4241a^2-5064ah+3600h^2}{34596b}$ |
| Social Welfare | $SW_1$ | $\frac{120a^2-59ah+339h^2}{578b}$ | < | $\frac{18971a^2-2208ah+62352h^2}{69192b}$ |
| | $SW_2$ | $\frac{(11a-10h)(57a+10h)}{2312b}$ | < | $\frac{2323a^2-2184ah-784h^2}{7688b}$ |
| | $SW_{total}$ | $\frac{1107a^2-696ah+1256h^2}{2312b}$ | < | $\frac{19939a^2-10932ah+27648h^2}{34596b}$ |

*Table 9 Retailer's monopoly vs. duopoly under dynamic contract when holding cost is high (a/12 <= h < a/4)*

| | | **Monopoly (Anand 2008)** | | **Duopoly** |
|---|---|---|---|---|
| Wholesale Price | $w_1$ | $\frac{1}{17}(9a-2h)$ | $>$ | $\frac{a}{2}$ |
| | $w_2$ | $\frac{2}{17}(3a+5h)$ | $<$ | $\frac{a}{2}$ |
| | $w_{avg}$ | $\frac{9a^2-4ah+8h^2}{19a-8h}$ | $<$ | $\frac{a}{2}$ |
| Retail Price | $p_1$ | $\frac{1}{17}(13a-h)$ | $>$ | $\frac{2a}{3}$ |
| | $p_2$ | $\frac{1}{34}(23a+10h)$ | $>$ | $\frac{2a}{3}$ |
| | $p_{avg}$ | $\frac{461a^2-84ah-104h^2}{34(19a-8h)}$ | $>$ | $\frac{2a}{3}$ |
| Consumer Purchase Qty | $Q_1$ | $\frac{4a+h}{17b}$ | $<$ | $\frac{a}{3b}$ |
| | $Q_2$ | $\frac{11a-10h}{34b}$ | $<$ | $\frac{a}{3b}$ |
| | $Q_{total}$ | $\frac{19a-8h}{34b}$ | $<$ | $\frac{2a}{3b}$ |
| Inventory | I | $\frac{5(a-4h)}{34b}$ | $>$ | 0 |
| Retailer (ea.) Sales | $q_1$ | $\frac{13a-18h}{34b}$ | $>$ | $\frac{a}{6b}$ |
| | $q_2+I$ | $\frac{3a+5h}{17b}$ | $>$ | $\frac{a}{6b}$ |
| | $q_{total}$ | $\frac{19a-8h}{34b}$ | $>$ | $\frac{a}{3b}$ |
| Retailer (ea.) Profit | $\pi R_1$ | $\frac{-13a^2+121ah+302h^2}{578b}$ | $<$ if $h<0.14969$; $>$ if $h>0.14969$ | $\frac{a^2}{36b}$ |
| | $\pi R_2$ | $\frac{181a^2-360ah-300h^2}{1156b}$ | $>$ | $\frac{a^2}{36b}$ |
| | $\pi R_{total}$ | $\frac{155a^2-118ah+304h^2}{1156b}$ | $>$ | $\frac{a^2}{18b}$ |
| Manufacturer Profit | $\pi M_1$ | $\frac{(13a-18h)(9a-2h)}{578b}$ | $>$ | $\frac{a^2}{6b}$ |
| | $\pi M_2$ | $\frac{2(3a+5h)^2}{289b}$ | $<$ | $\frac{a^2}{6b}$ |
| | $\pi M_{total}$ | $\frac{9a^2-4ah+8h^2}{34b}$ | $<$ | $\frac{a^2}{3b}$ |
| SC Profit | $\pi_1$ | $\frac{104a^2-67ah+338h^2}{578b}$ | $<$ | $\frac{2a^2}{9b}$ |
| | $\pi_2$ | $\frac{(11a-10h)(23a+10h)}{1156b}$ | $<$ | $\frac{2a^2}{9b}$ |
| | $\pi_{total}$ | $\frac{461a^2-254ah+576h^2}{1156b}$ | $<$ | $\frac{4a^2}{9b}$ |
| Consumer Surplus | $CS_1$ | $\frac{(4a+h)^2}{578b}$ | < | $\frac{a^2}{18b}$ |
| | $CS_2$ | $\frac{(11a-10h)^2}{2312b}$ | < | $\frac{a^2}{18b}$ |
| | $CS_{total}$ | $\frac{185a^2-188ah+104h^2}{2312b}$ | < | $\frac{a^2}{9b}$ |
| Social Welfare | $SW_1$ | $\frac{120a^2-59ah+339h^2}{578b}$ | < | $\frac{5a^2}{18b}$ |
| | $SW_2$ | $\frac{(11a-10h)(57a+10h)}{2312b}$ | < | $\frac{5a^2}{18b}$ |
| | $SW_{total}$ | $\frac{1107a^2-696ah+1256h^2}{2312b}$ | < | $\frac{5a^2}{9b}$ |

*Table 10 Retailer's monopoly vs. duopoly under commitment contract*

| | | **Monopoly (Anand 2008)** | | **Duopoly** |
|---|---|---|---|---|
| Wholesale Price | $w_1$ | $\frac{a}{2}$ | = | $\frac{a}{2}$ |
| | $w_2$ | $\frac{a}{2}$ | = | $\frac{a}{2}$ |
| | $w_{avg}$ | $\frac{a}{2}$ | = | $\frac{a}{2}$ |
| Retail Price | $p_1$ | $\frac{3a}{4}$ | > | $\frac{2a}{3}$ |
| | $p_2$ | $\frac{3a}{4}$ | > | $\frac{2a}{3}$ |
| | $p_{avg}$ | $\frac{3a}{4}$ | > | $\frac{2a}{3}$ |
| Consumer Purchase Qty | $Q_1$ | $\frac{a}{4b}$ | < | $\frac{a}{3b}$ |
| | $Q_2$ | $\frac{a}{4b}$ | < | $\frac{a}{3b}$ |
| | $Q_{total}$ | $\frac{a}{2b}$ | < | $\frac{2a}{3b}$ |
| Inventory | I | 0 | = | 0 |
| Retailer (ea.) Sales | $q_1$ | $\frac{a}{4b}$ | > | $\frac{a}{6b}$ |
| | $q_2 + I$ | $\frac{a}{4b}$ | > | $\frac{a}{6b}$ |
| | $q_{total}$ | $\frac{a}{2b}$ | > | $\frac{a}{3b}$ |
| Retailer (ea.) Profit | $\pi R_1$ | $\frac{a^2}{16b}$ | > | $\frac{a^2}{36b}$ |
| | $\pi R_2$ | $\frac{a^2}{16b}$ | > | $\frac{a^2}{36b}$ |
| | $\pi R_{total}$ | $\frac{a^2}{8b}$ | > | $\frac{a^2}{18b}$ |
| Manufacturer Profit | $\pi M_1$ | $\frac{a^2}{8b}$ | < | $\frac{a^2}{6b}$ |
| | $\pi M_2$ | $\frac{a^2}{8b}$ | < | $\frac{a^2}{6b}$ |
| | $\pi M_{total}$ | $\frac{a^2}{4b}$ | < | $\frac{a^2}{3b}$ |
| SC Profit | $\pi_1$ | $\frac{3a^2}{16b}$ | < | $\frac{2a^2}{9b}$ |
| | $\pi_2$ | $\frac{3a^2}{16b}$ | < | $\frac{2a^2}{9b}$ |
| | $\pi_{total}$ | $\frac{3a^2}{8b}$ | < | $\frac{4a^2}{9b}$ |
| Consumer Surplus | $CS_1$ | $\frac{a^2}{32b}$ | < | $\frac{a^2}{18b}$ |
| | $CS_2$ | $\frac{a^2}{32b}$ | < | $\frac{a^2}{18b}$ |
| | $CS_{total}$ | $\frac{a^2}{16b}$ | < | $\frac{a^2}{9b}$ |
| Social Welfare | $SW_1$ | $\frac{7a^2}{32b}$ | < | $\frac{5a^2}{18b}$ |
| | $SW_2$ | $\frac{7a^2}{32b}$ | < | $\frac{5a^2}{18b}$ |
| | $SW_{total}$ | $\frac{7a^2}{16b}$ | < | $\frac{5a^2}{9b}$ |